\documentclass[useAMS,usenatbib]{mn2e}

\usepackage[T1]{fontenc}
\usepackage{pslatex}
\usepackage{amsmath}
\usepackage{amsfonts}
\usepackage{color}
\usepackage{url}
\usepackage{graphicx}
\usepackage{subfigure}
\usepackage{amssymb}
\usepackage{textcomp}
\usepackage{soul}
\usepackage{booktabs}
\usepackage{array}
\usepackage{hyperref}
\usepackage{enumerate}
\usepackage{bm}
 \graphicspath{{images/}}

{\newif\ifnotend
\notendtrue
\def\veclist{ABCDEFGHIJKLMNOPQRSTUVWXYZabcdefghijklmnopqrstuvwxyz.}
\def\top#1#2.{#1}
\def\tail#1#2.{#2.}
\loop\expandafter\xdef\csname v\expandafter\top\veclist\endcsname%
{{\noexpand\bf\expandafter\top\veclist}}
\edef\veclist{\expandafter\tail\veclist}
\if\veclist.\notendfalse\fi\ifnotend\repeat}
\def\Msunyr{\, \rm M_\odot \, yr^{-1}}

\def\pa{\partial}

\mathchardef\mhyphen="2D

\title[Do nuclear rings form at the shear minimum?]{Do nuclear rings in barred galaxies form at the shear minimum of the rotation curve?}
\author[Sormani \& Li]{Mattia C. Sormani$^{1}$ and Zhi Li$^2$\\
$^1$ Universit\"{a}t Heidelberg, Zentrum f\"{u}r Astronomie, Institut f\"{u}r theoretische Astrophysik, Albert-Ueberle-Str. 2, 69120 Heidelberg, Germany; \\
$^2$ Tsung-Dao Lee Institute, Shanghai Jiao Tong University, 800 Dongchuan Road, Shanghai 200240, P.R China; email: zli0804@sjtu.edu.cn
}

\begin{document}

\date{}

\def\p{\partial}
\def\Omegap{\Omega_{\rm p}}

\newcommand{\di}{\mathrm{d}}
\newcommand{\bfx}{\mathbf{x}}
\newcommand{\bfe}{\mathbf{e}}
\newcommand{\vlos}{\mathrm{v}_{\rm los}}
\newcommand{\Tspin}{T_{\rm s}}
\newcommand{\Tb}{T_{\rm b}}
\newcommand{\degree}{\ensuremath{^\circ}}
\newcommand{\Th}{T_{\rm h}}
\newcommand{\Tc}{T_{\rm c}}
\newcommand{\bfr}{\mathbf{r}}
\newcommand{\bfv}{\mathbf{v}}
\newcommand{\pc}{\,{\rm pc}}
\newcommand{\kpc}{\,{\rm kpc}}
\newcommand{\Myr}{\,{\rm Myr}}
\newcommand{\Gyr}{\,{\rm Gyr}}
\newcommand{\kms}{\,{\rm km\, s^{-1}}}
\newcommand{\de}[2]{\frac{\partial #1}{\partial {#2}}}
\newcommand{\cs}{c_{\rm s}}
\newcommand{\rb}{r_{\rm b}}
\newcommand{\rqu}{r_{\rm q}}
\newcommand{\nuP}{\nu_{\rm P}}
\newcommand{\thetaobs}{\theta_{\rm obs}}
\newcommand{\hatn}{\hat{\textbf{n}}}
\newcommand{\hatx}{\hat{\textbf{x}}}
\newcommand{\haty}{\hat{\textbf{y}}}
\newcommand{\hatz}{\hat{\textbf{z}}}
\newcommand{\hatX}{\hat{\textbf{X}}}
\newcommand{\hatY}{\hat{\textbf{Y}}}
\newcommand{\hatZ}{\hat{\textbf{Z}}}
\newcommand{\hatN}{\hat{\textbf{N}}}

\maketitle

\begin{abstract}
It has been recently suggested that (i) nuclear rings in barred galaxies (including our own Milky Way) form at the radius where the shear parameter of the rotation curve reaches a minimum; (ii) the acoustic instability of Montenegro et al. is responsible for driving the turbulence and angular momentum transport in the central regions of barred galaxies. Here we test these suggestions by running simple hydrodynamical simulations in a logarithmic barred potential. Since the rotation curve of this potential is scale-free, the shear minimum theory predicts that no ring should form. We find that in contrast to this prediction, a ring does form in the simulation, with morphology consistent with that of nuclear rings in real barred galaxies. This proves that the presence of a shear-minimum is not a necessary condition for the formation of a ring. We also find that perturbations that are predicted to be acoustically unstable wind up and eventually propagate off to infinity, so that the system is actually stable. We conclude that (i) the shear-minimum theory is an unlikely mechanism for the formation of nuclear rings in barred galaxies; (ii) the acoustic instability is a spurious result and may not be able to drive turbulence in the interstellar medium, at least for the case without self-gravity. The question of the role of turbulent viscosity remains open.
\end{abstract}


\section{Introduction} \label{sec:intro}

Nuclear gaseous rings are a common feature of barred galaxies \citep{Comeron2010}. It has recently attracted attention a theory proposed by \cite{KrumholzKruijssen2015} and \cite{KrumholzKruijssen2017}, and originally by \citealt{Lesch+1990}, according to which nuclear rings form at the radius where the shear due to differential rotation reaches a minimum. 

In the \cite{KrumholzKruijssen2015} version of this theory, angular momentum transport is driven by the acoustic instability of \cite{Montenegro+99}. Transport is expected to be more efficient where shear is higher, so that gas is expected to pile up and form a ring where the transport becomes less efficient, i.e. at the radius of minimum shear. Moreover, according to \cite{Montenegro+99} and \cite{KrumholzKruijssen2015} acoustic instabilities are driven by pressure rather than gravity, and as a consequence they are predicted to occur even in the non-self gravitating case. 

As noted by \cite{KrumholzKruijssen2015}, hydrodynamical simulations of gas flow in barred potentials should in principle be able to test predictions of their theory. The goal of this short paper is to perform such tests.

\section{The test}

The test consists in running hydrodynamical simulations in the logarithmic barred potential:
\begin{equation} \label{eq:logphi}
\Phi(x,y) = \frac{v_0^2}{2} \log \left( x^2 + \frac{y^2}{q^2} \right)\, , 
\end{equation}
where $q \leq 1$. The potential is assumed to be rigidly rotating with pattern speed $\Omegap$. In this potential, all the multipoles, and in particular the rotation curve $v_{\rm c}(R) = v_0$, are constant (see Appendix \ref{sec:logphi}), so there is no shear minimum. Therefore, according to the shear-minimum theory, no ring should form in such a potential. We use our simulations to test this prediction. Moreover, the dispersion relation of \cite{Montenegro+99} predicts that a non-self gravitating gaseous disc flowing in this potential should be unstable to non-axisymmetric modes (see \S\ref{sec:acoustic}). Therefore, this potential also allows us to test whether the acoustic instability is able to drive turbulence in the interstellar medium (ISM).

\subsection{Equations of motion}
We assume the gas to be isothermal
\begin{equation}
P = \cs^2 \rho
\end{equation}
where $\cs = {\rm constant}$. We neglect the gas self-gravity and its associated additional physics. The equations of motion in the rotating frame co-rotating with the bar at ${\bm \Omegap} = \Omegap \hat{\mathbf{e}}_z$ are the continuity and Euler equations:
\begin{align}
\pa_t \rho + \nabla\cdot \left( \rho \bfv \right) &= 0 \label{eq:continuity} \\
\pa_t \bfv + \left(\bfv \cdot \nabla \right) &= - \cs^2 \frac{\nabla \rho}{\rho} - \nabla\Phi - 2 {\bm \Omegap} \times \bfv + \Omegap^2 R \, \hat{\mathbf{e}}_R \label{eq:euler}
\end{align}
where $\rho$ is the surface density, $\bfv$ is the velocity, $(R,\theta,z)$ denote standard cylindrical coordinates, $\hat{\mathbf{e}}_R$ is the unit vector in the radial direction and $\hat{\mathbf{e}}_z$ the unit vector in the $z$ direction. 

\subsection{Numerical setup} 

To solve Eqs. \eqref{eq:continuity} and \eqref{eq:euler}, we use the public grid code {\sc Pluto} \cite{Mignone+2007} version 4.3. We use a two-dimensional static polar grid in the region $R \times \theta = [0.005,10] \kpc \times [0, 2\pi]$. The grid is logarithmically spaced in $R$ and uniformly spaced in $\theta$ with $4096 \times 4096$ cells. The resolution along the $R$ direction is approximately $\Delta R=0.00186~R$, i.e. we have a resolution of $\Delta R=0.186\pc$ at $R=100\pc$. The acoustic instability, that in the model of \cite{KrumholzKruijssen2015} is postulated to drive the angular momentum transport, is thus very well resolved (see \S\ref{sec:acoustic}). We use the following parameters: {\sc rk2} time-stepping, no dimensional splitting, {\sc hll} Riemann solver and the default flux limiter. We solve the equations in the frame rotating at $\Omegap$ by using the {\sc rotating\_frame = yes} switch. Boundary conditions are reflective on the inner boundary at $R=0.005\kpc$ and outflow on the outer boundary at $R=10.0\kpc$.

The initial density distribution is taken to be uniform with value $\rho_0$. The density units are arbitrary since the equations of motion \eqref{eq:continuity} and \eqref{eq:euler} are invariant under density rescaling, so without loss of generality we set $\rho_0=1$. In order to avoid transients, we introduce the bar gradually, as is common practice in this type of simulations \citep[e.g.][]{Athan92b}. We start with gas in equilibrium on circular orbits in an axisymmetric logarithmic potential with $q=1$ and then we turn on the non-axisymmetric part of the potential linearly during the first $100\Myr$ in such a way that the monopole of the potential is kept constant (see Eq. \ref{eq:monopole}), until the final value $q=0.9$ is reached. The other parameters of the simulation are listed in Table \ref{tab:1}.

We have checked that our results do not depend on the particular type of code that we are using by testing the same simulation setup of gas flow in external barred potentials using the SPH code {\sc Phantom} \citep[][]{price18}. Reassuringly, this yields results very similar to {\sc Pluto}.

\begin{table}
\centering
\begin{tabular}{lr}  
\toprule
Parameter    	& Value 				\\
\midrule
q			& $0.9$ or $1.0$				\\
$v_0$      		& $200 \kms$			\\
$\cs$	  	& $10 \kms$			\\
$\Omegap$	& $40 \kms \kpc^{-1}$	\\
\bottomrule
\end{tabular}
\caption{Parameters of the simulations.} \label{tab:1}
\end{table}

\section{Result} \label{sec:results}

\begin{figure*}
\centering
\includegraphics[width=1.0\textwidth]{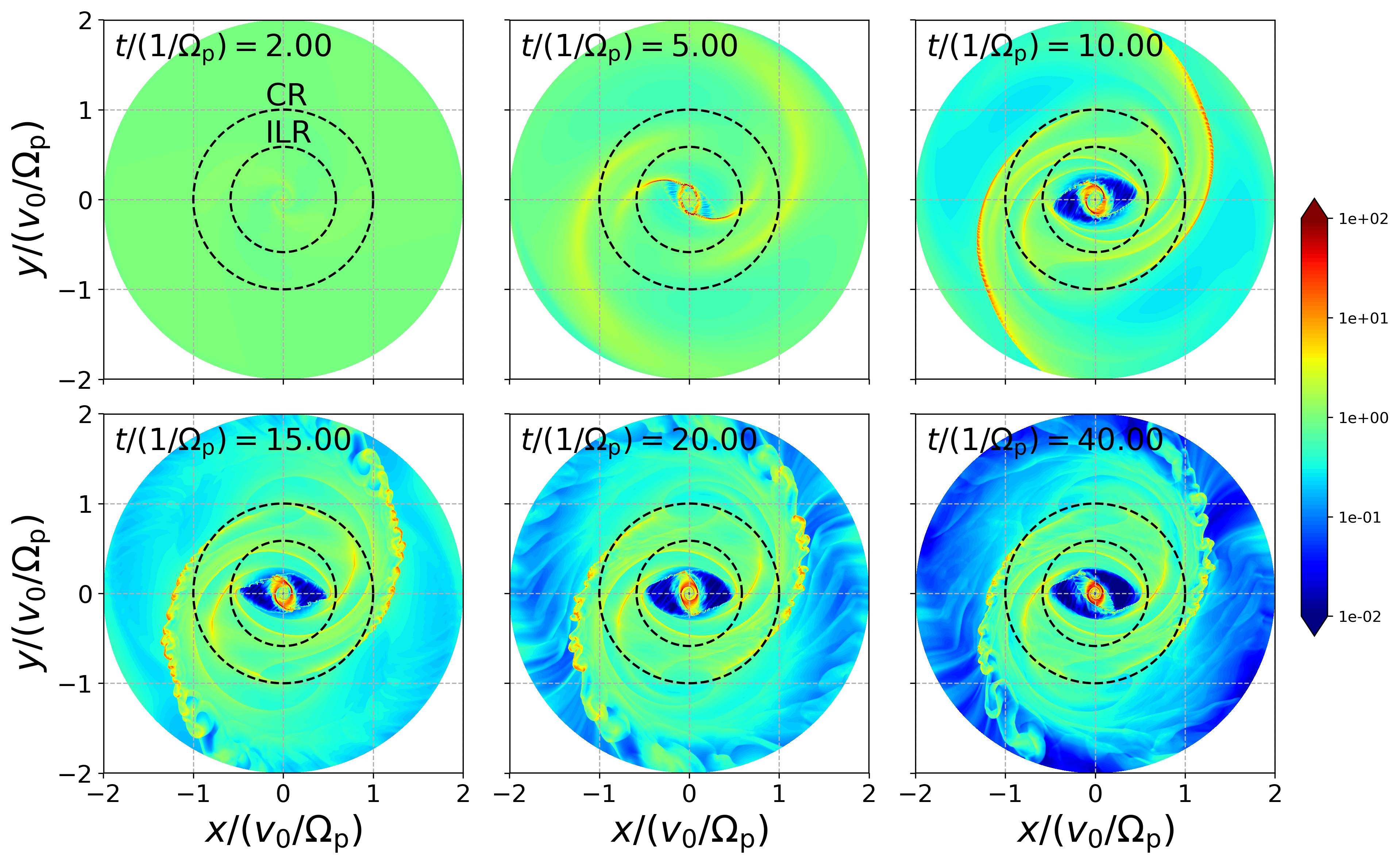}
\caption{Surface density of the simulation for the parameter listed in Table \ref{tab:1}. A long-lasting nuclear gaseous ring forms at $R\simeq0.1 (v_0/\Omegap)$, in contrast to the prediction of the shear-minimum theory. The black dashed circles indicate the inner Lindblad and corotation resonances (see Eqs. \ref{eq:ILR}-\ref{eq:CR}). Note the range of the colorbar is $0.01 \mhyphen 100$, and the initial uniform density is set to $1$.}
\label{fig:1}
\end{figure*}

The result of the simulation with $q=0.9$ is shown in Fig. \ref{fig:1}. A nuclear ring forms at $R\simeq0.1 (v_0/\Omegap)$, in contrast to the prediction of the shear-minimum theory. The morphology of the ring is consistent with that of typical simulations of gas flow in barred potentials \citep[e.g.][]{Athan92b,Kim++2012a,Kim++2012b,SBM2015a,Li+2015} and with that of real barred galaxies \citep[e.g.][]{Sellwood1993,Comeron2010}. This proves that the presence of a shear minimum is not a necessary condition for the formation of a ring.

\section{Discussion}

\subsection{Dimensional analysis} \label{sec:dimensional}

How can a ring with a definite size form if the potential given by Eq. \eqref{eq:logphi} is scale-free? The answer is that, in the case $q \neq 1$, the scale invariance is broken by the fact that the bar potential rotates with pattern speed $\Omegap$. This sets a dimensional scale in the problem, $L = v_0 / \Omegap$. Therefore, the size of any feature that forms in the simulation, including rings, must scale with $L$. Note that in contrast, in the shear-minimum theory, $\Omegap$ plays no role in setting the radius of the ring.

It is useful to do a more complete dimensional analysis. The problem has four parameters in total:
\begin{equation}
v_0, \qquad	\Omegap, \qquad q, \qquad \cs\,.
\end{equation}
Without loss of generality we can set $v_0=\Omegap=1$, since this amounts to choosing the spatial and temporal units of the problem. This leaves us with two dimensionless parameters:
\begin{equation}
\alpha \equiv \cs/v_0, \qquad \beta \equiv 1-q\,.
\end{equation}
The size of the ring must depend only on these two quantities. Hence, from dimensional analysis only, we obtain that the size of the ring must be a function of the following type:
\begin{equation}
R_{\rm ring}(\alpha,\beta) = R_0\, F(\alpha,\beta)
\end{equation}
where $F=F(\alpha,\beta)$ is a dimensionless factor, and $R_0$ is the radius of the ring in the limit of vanishing sound speed ($\alpha\to0$) and vanishing bar strength ($\beta\to0$), so that $F(0,0)=1$. The linear analysis of \cite{GoldreichTremaine1979} shows that in this limit the bar clears a small gap at the radius of the inner Lindblad resonance (ILR), and gas accumulates just inside this radius. Thus we must have $R_0=R_{\rm ILR}$. Numerical simulations \citep{PatsisAthanassoula2000,Kim++2012a,Kim++2012b,SBM2015a,Li+2015} show that $R_{\rm ring}$ always decreases when sound speed and/or bar strength are increased. Thus in summary the size of the ring must be
\begin{equation} \label{eq:Rring}
R_{\rm ring} = R_{\rm ILR} F(\alpha,\beta)
\end{equation}
with
\begin{equation}
F(\alpha,\beta) \leq 1
\end{equation}
with $F$ being a decreasing function of $\alpha$ and $\beta$. The exact form of $F$ is complicate and with our current understanding it can be determined only numerically. 

The above discussion clearly shows that there exist at least another mechanism which is (a) physically distinct from the shear minimum theory, and (b) capable of forming rings with characteristics similar to those observed in the nuclear rings in barred galaxies. Equation \eqref{eq:Rring}, which follows from very general considerations, expresses the size of the ring formed by this mechanism. Note that according to this equation the size of the ring depends on $\Omegap$ (through $R_{\rm ILR}$), $\cs$ and $q$. This is at variance with the shear-minimum theory, according to which the size of the ring does not depend on $\Omegap$, nor on $\cs$, nor on $q$.

The gas density distribution in Fig. \ref{fig:1} has all the typical morphological characteristics (low density of gas in and around the bar region, narrow lanes which are the loci of shocks and which spiral into the nuclear ring) that are generally found in more realistic simulations of barred galaxies \citep[e.g.][]{Sormani2018} and in observations of real barred galaxies \citep[e.g.][]{Sellwood1993}. While we acknowledge that this could be a coincidence, we find it more likely that the same mechanism at work in our simulation is at work also in real barred galaxies. In alternative, one must then explain why the mechanism at work in our simulation is not relevant for real galaxies despite the gas flow and morphology similarity.

We therefore conclude that (i) the shear minimum theory is not responsible for the formation of the ring in the simulation in Fig. \ref{fig:1}; (ii) there exist at least another mechanism which is physically distinct from the shear minimum theory one and which is capable of forming rings with characteristics consistent with those observed in the nuclear rings in barred galaxies.

\subsection{The acoustic instability} \label{sec:acoustic}

The shear-minimum theory presented by \cite{KrumholzKruijssen2015} relies on the acoustic instability reported by \cite{Montenegro+99} as the source of angular momentum transport. We therefore want to make sure that in our simulation we are able to resolve the possible appearance of this instability and analyse its consequences.

\cite{Montenegro+99} consider an axisymmetric, barotropic ($P=P(\rho)$), differentially rotating fluid disc and study the propagation of small (linear) axisymmetric and non-axisymmetric perturbations on top of it, similarly to the classical analysis of \cite{LinShu1964}. They use a \emph{local} WKB approximation, but compared to Lin-Shu they keep one extra order in the quantity $1/|kR|$ (assumed small in the WKB approximation), where $k$ is the radial wavenumber of the perturbation. As a result, their dispersion relation (their equation 10) contains one extra term compared to the well-known Lin-Shu dispersion relation. This difference is relevant only in the non-axisymmetric ($m\neq0$) case, while in the axisymmetric case ($m=0$) their dispersion relation reduces to the standard Lin-Shu dispersion relation (which in the axisymmetric case gives the \cite{Toomre1964} criterion). The \cite{Montenegro+99} analysis applies to both self-gravitating and non self-gravitating discs. They report that even in the absence of self-gravity, the resulting dispersion relation admits non-axisymmetric unstable modes, which they dubbed "acoustic" because they are driven by pressure and shear rather than by gravity.

For the non self-gravitating case studied in this paper, the dispersion relation is given by eq. (15) of \cite{Montenegro+99}: 
\begin{equation} \label{eq:disprel}
\nu^4- \left(2+\frac{\cs^2 k^2}{\kappa^2}\right)\nu^2+\frac{\cs^2 \left( k^2 + 2 m^2 / R^2 \right)}{\kappa^2}+1=0,
\end{equation}
where the dimensionless frequency $\nu$ is defined as $\nu=(\omega-m\Omega)/\kappa$, $\kappa=\kappa(R)$ is the epicyclic frequency (see Eq. \ref{eq:kappa}), and $\omega$ is the frequency of the perturbation. Modes with $\operatorname{Im}(\omega)>0$ grow exponentially in time, and are formally unstable. Modes with $\operatorname{Im}(\omega)=0$ or $\operatorname{Im}(\omega)<0$ are oscillating and decaying solutions.

The condition for acoustic instability to occur is given by eq. (16) of \cite{Montenegro+99}:
\begin{equation}\label{eq:stabcond}
    \left(k^2 + \frac{m^2}{R^2}\right) < \frac{4 \Omega m}{\cs R},
\end{equation}
where $\Omega=\Omega(R)$ is the angular frequency derived from the rotation curve, and $m$ is an integer ($m=0$ for axisymmetric perturbations). We see from Eq.~\eqref{eq:stabcond} that only non-axisymmetric modes can be unstable. Inserting the rotation curve of the logarithmic potential (Eq. \ref{eq:Omega}), our values of the parameters (Table \ref{tab:1}), and setting $m=2$ as in \cite{KrumholzKruijssen2015}, this condition becomes
\begin{equation} \label{eq:cond}
    k^2  < \frac{156}{R^2},
\end{equation}
which means that instability occurs for wavelength larger than $\lambda = ({2 \pi}/{156})^{1/2} R \simeq 0.2 R$. For example, at $R=1\kpc$ the smallest unstable wavelength is $\lambda = 200\pc$. These wavelengths are well resolved in our simulation. 

Contrary to the predictions of \cite{Montenegro+99}, we find that the acoustic instability which forms the basis of the \cite{KrumholzKruijssen2015} theory does not play a role in our simulation. The turbulence visible in Fig. \ref{fig:1} is caused by the wiggle instability, not by the acoustic instability (see \citealt{WadaKoda2004,KimKimKim2014,Sormani+2017}). We have performed various tests in order to confirm that acoustic instabilities do not occur.

The dispersion relation of \cite{Montenegro+99} is derived for an axisymmetric background potential, not for a barred potential as in Fig. \ref{fig:1}. Therefore, a cleaner test of the acoustic instability can be performed by running simulations similar to the one in Fig. \ref{fig:1} but with $q=1$ (i.e., axisymmetric potential). In this way, the background axisymmetric steady state on top of which perturbations propagate are exactly the same as in the \cite{Montenegro+99} analysis. The equations of motions (Eqs. \ref{eq:continuity}-\ref{eq:euler}) are also in principle the same, with the difference that \cite{Montenegro+99} performs various approximation in their analysis (linearity and the WKB approximation), while our code solves the equations exactly.

The first test is to run a simulation identical to the one shown in Fig. \ref{fig:1} but with $q=1$. This system is predicted to have unstable modes by \cite{Montenegro+99}. We find that in contrast to this prediction, the disc in the simulation is remarkably stable, to the point that after several rotations ($t \Omegap = 40$) the density distribution still looks identical to the uniform initial conditions. 

\begin{figure*}
\centering
\includegraphics[width=1.0\textwidth]{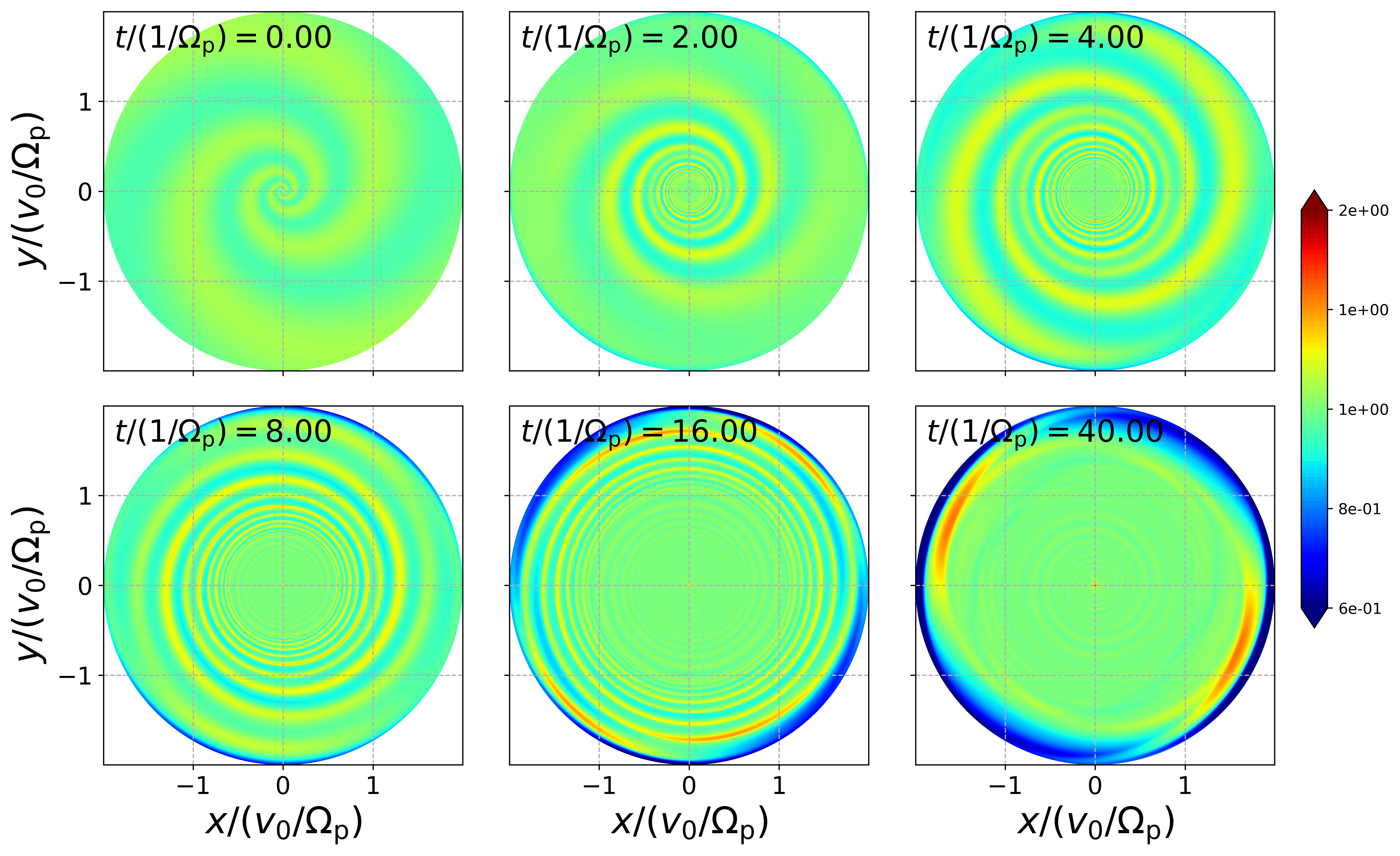}
\caption{Evolution of a spiral mode with $k=6/R$ which is acoustically unstable according to the dispersion relation \citet{Montenegro+99} (Eq.~\ref{eq:disprel}). Contrary to the predictions of the latter, the mode winds up and propagates off to infinity, leaving behind a uniform disc. Hence, the fluid disc is actually stable. Shown is the surface density distribution. Note the range of the colorbar is $0.6 \mhyphen 2$, and the initial uniform density is set to $1$.}
\label{fig:2}
\end{figure*}

We now consider the possibility that the reason why the instability does not show up in the previous test is because the initial conditions are too smooth, so that unstable perturbations do not have an initial seed from which to grow. In order to test whether this is the case, we perform a second test, in which we start by imposing an $m=2$ spiral perturbation with wave number $k=6/R$, which is predicted to be unstable according to Eq.~\eqref{eq:cond}. This spiral perturbation is included on top of the background steady state (i.e. $\rho_0=1$,$v_\theta=v_0$, and $v_R=0$) in the initial conditions as
\begin{align}
\rho_1(R,\theta)   & 	=  \operatorname{Re}\left[\exp(2i\theta)\rho_{\rm a}(R) \right]\\
v_R(R,\theta)      & 	=   \operatorname{Re}\left[\exp(2i\theta)v_{\rm Ra}(R)\right]\\
v_\theta(R,\theta) & 	=  \operatorname{Re}\left[\exp(2i\theta)v_{\rm \theta a}(R)\right],
\end{align}
where $\rho_{\rm a}(R)=0.05 \exp(6i \log R)$, $v_{\rm Ra}(R)=i/\Delta[(\omega-2\Omega)\di h_{\rm a}/ \di R - 4h_{\rm a}\Omega/R$], and $v_{\rm \theta a}(R)=-1/\Delta[2B \di h_{\rm a}/ \di R+2h_{\rm a}(\omega-2\Omega)/R]$. The other variables are defined as $\Delta=\kappa^2-(\omega-2\Omega)^2$, $B=-\kappa^2/4\Omega$, and $h_{\rm a}(R)= \cs^2\rho_{\rm a}(R)/\rho_0$. These equations are basically equations (6.42)-(6.45) of \citet{BT2008}. Note that the growth rate $\omega$ of the unstable mode $k=6/R$ is obtained by solving Eq.~\eqref{eq:disprel}, which yields two possible unstable frequencies, one with $\operatorname{Re}(\nu)>0$ and one with $\operatorname{Re}(\nu)<0$. We have tested both values. The amplitude of the perturbation spiral surface density is $5\%$ of the initial uniform density, and the perturbed velocities on radial and azimuthal directions are of order of $1\kms$. Figure~\ref{fig:2} shows the gas surface density evolution with the initial condition above and choosing the root with $\operatorname{Re}(\nu)>0$. We find that the modes wind up and eventually propagate off to infinity, leaving behind a smooth disc, so that the system is actually stable. When choosing $\operatorname{Re}(\nu)<0$, results are similar except that the wave first propagates inwards, and when it reaches the inner boundary it either bounces back and propagates off to infinity or gets absorbed depending on the boundary conditions. We have also tested the case with $k=12/R$, and the results are similar.

We note that also the \cite{LinShu1964} dispersion relation formally admits non-axisymmetric growing modes when the Toomre parameter is less than unity ($Q<1$). However, this is known to be a spurious result: \cite{GoldreichLyndenbell1965} and \cite{JulianToomre1966} have shown that non-axisymmetric disturbances in a fluid disc actually wind up and propagate off to infinity (see also footnote 6 at pg. 494 of \citealt{BT2008}). This is exactly the behaviour we see in our simulations shown in Fig. \ref{fig:2}. This is in hindsight not surprising, since as discussed above the \cite{Montenegro+99} is just a refinement of the \cite{LinShu1964} dispersion relation. We conclude that the acoustic instability is a spurious results and cannot drive turbulence in the central regions of barred galaxies.

\subsection{What is the role of viscosity?} \label{sec:viscosity}

The results of the previous section indicate that the acoustic instability cannot drive the turbulence. However, the gas in the central regions of barred galaxies \emph{is} observed to be turbulent. This should generate turbulent viscosity and hence angular momentum transport regardless of the driving mechanism. Using considerations from kinetic theory applied to an ensemble of molecular clouds, \cite{LyndenBellPringle1974} estimated the turbulent viscosity coefficient in the central region of galaxies as
\begin{equation} \label{eq:nuism}
\nu\approx\frac{1}{3}c\frac{a}{f} \simeq 5 \times 10^2 \pc^2 \Myr^{-1} 
\end{equation}
where $a\sim 10 \pc$ is the average cloud size, $c \sim 15 \kms$ is the cloud velocity dispersion and $f\sim 0.1$ is the clouds volume filling factor. The typical timescale associated with this viscosity coefficient is $R^2 / \nu$, where $R$ is the radius of the nuclear ring. For $R=200 \pc$ we get a typical timescale of $80\Myr$, which is relatively short. This suggests that turbulent viscosity should significantly affect the dynamics of the gas. Thus, it is possible turbulent angular momentum transport plays a role in the central regions of barred galaxies, as in the shear minimum theory. However the situation in complicated by several facts: 
\begin{enumerate}
\item The presence of non-axisymmetric motions, as in the centre of barred galaxies, can create regions of ``reverse shear'' in which the sign of the angular momentum flux is reversed (i.e. it is the opposite of a standard axisymmetric accretion disc), with dramatic consequences for the mass transport \citep{Sormani+2018}.
\item The mean free path of molecular clouds in a nuclear ring can be estimated as $\lambda_{\rm cloud} = 1/{\sigma n}$, where $\sigma$ is the cross section and $n$ is the volume number density of the clouds. Using $\sigma = \pi a^2$ and that (assuming roughly spherical clouds) the volume filling factor can be written as $f = n (4 \pi a^3/3)$, we have
\begin{equation}
\lambda_{\rm cloud} =  \frac{4a}{3f} \simeq 130 \pc,
\end{equation} 
which is comparable to the typical width of nuclear rings. Hence, the fluid approximation on which the turbulent viscosity picture is based breaks down.
\item The presence of galactic winds and/or AGN feedback may remove the gas from the ring on a timescale shorter than the timescale necessary for the viscosity to act. Indeed, galactic winds can remove mass at a rate of several $\Msunyr$ \citep[e.g.][]{Chisholm+2017}. For example, in the Milky Way, which has a currently quiescent supermassive black hole (SgrA*), observations suggest that the outflow associated with the Fermi Bubbles \citep{Su+2010} removes the gas at a rate of $\sim 1 \Msunyr$, although with large uncertanties (see \citealt{Bordoloi+2017,Diteodoro+2018} and discussion in Sect. 4.2 of \citealt{SormaniBarnes2019}). At this rate, the whole central molecular zone, which has a total gas mass of $\sim 5 \times 10^7 \, \rm M_\odot$ \citep{Dahmen+1998}, would take only 50 Myr to disappear, a time shorter than the viscous timescale estimated above.
\end{enumerate}

We conclude that while it is possible that turbulent viscosity significantly affects the dynamics of nuclear rings, its exact role is an open question. Addressing this question requires further numerical experiments that take into account the fact that the ISM is a highly compressible turbulent medium, for example simulations similar to the present one with the addition of controlled forced turbulent viscosity. However, the influence of highly compressible turbulence on the dynamics of the ISM, and how mixing and transport processes can be captured by macroscopic prescriptions such as a diffusion coefficient, are complex and specialised research areas on their own \cite[e.g.][]{KlessenLin2003,SchmidtFederrath2011}. Thus, while this is a worthwhile direction of future investigation, it lies outside of the scope of the current paper.

\section{Conclusion}

We can summarise the findings of this paper as follows:
\begin{enumerate}
\item The presence of a shear-minimum is not a necessary condition for the formation of nuclear rings (\S\ref{sec:results}).
\item There is at least another physical mechanism which is physically distinct from the minimum-shear and is capable of generating nuclear rings with morphologies consistent with those of real barred galaxies. Dimensional analysis and general considerations show that this is the mechanism that generates the ring in Fig. \ref{fig:1} (\S\ref{sec:dimensional}).
\item The acoustic instability is a spurious result and cannot drive turbulence in the ISM (\S\ref{sec:acoustic}).
\item Order of magnitude estimates suggest that turbulent viscosity should be able to significantly affect the dynamics of the gas. However, its precise role is an open question that deserves further investigation (\S\ref{sec:viscosity}).
\end{enumerate}

\section*{Acknowledgements}

We are grateful to Ralf Klessen, Mark Krumholz, Jerry Sellwood and Scott Tremaine for helpful discussions. MCS acknowledges financial support from the German Research Foundation (DFG) via the collaborative research center (SFB 881, Project-ID 138713538) ``The Milky Way System'' (subprojects A1, B1, B2, and B8). 
This work made use of the facilities of the Center for High Performance Computing at
Shanghai Astronomical Observatory.  

\def\aap{A\&A}\def\aj{AJ}\def\apj{ApJ}\def\mnras{MNRAS}\def\araa{ARA\&A}\def\aapr{Astronomy \&
 Astrophysics Review}\def\apjs{ApJS}\def\apjl{ApJ}\def\pasj{PASJ}\def\nat{Nature}\def\prd{Phys. Rev. D}
\def\ssr{Space Sci. Rev.}\def\pasp{PASP}\def\aaps{A\&AS}\def\pasa{Publications of the Astron. Soc. of Australia}
\def\pre{Phys. Rev. E}

\bibliographystyle{mn2e}
\bibliography{bibliography}

\appendix

\section{Properties of the logarithmic potential} \label{sec:logphi}
Substituting polar coordinates $x = R \cos\theta$, $y=R \sin\theta$ into \eqref{eq:logphi} and expanding, we obtain the multipole expansion of the logarithmic potential
\begin{equation} \label{eq:multipoles}
\Phi(R,\theta) = v_0^2 \left[ \log\left(\frac{(1+q)^2}{4q} R \right) - \sum_{n=1}^{\infty} \frac{1}{n}\left[ \frac{1-q}{1+q}\right]^n \cos(2n\theta)  \right] \,.
\end{equation}
The monopole is
\begin{equation} \label{eq:monopole}
\Phi_0(R) = v_0^2 \log\left(\frac{(1+q)^2}{4q} R \right)
\end{equation}
From the monopole we find that the rotation curve is constant
\begin{equation}
v_{\rm c}(R) = \sqrt{ R \frac{\di \Phi_0}{\di R}} = v_0 \,.
\end{equation}
The angular velocity is
\begin{equation} \label{eq:Omega}
    \Omega(R) = \frac{v_0}{R},
\end{equation}
the epicyclic frequency is
\begin{equation} \label{eq:kappa}
\kappa = \sqrt{ \frac{2 \Omega }{R} \frac{\di}{\di R} \left( R^2 \Omega \right)} = \sqrt{2} \frac{v_0}{R} \,,
\end{equation}
and the radii of the inner Lindblad resonance (ILR), outer Lindblad resonance (OLR) and corotation resonance (CR) are:
\begin{align}
R_{\rm ILR} 	& = (2 - \sqrt{2}) \frac{v_0}{\Omegap} \simeq  0.59 \frac{v_0}{\Omegap} \,, \label{eq:ILR} \\ 
R_{\rm CR} 	& = \frac{v_0}{\Omegap}	\,, \label{eq:CR} \\
R_{\rm OLR} 	& = (2 + \sqrt{2}) \frac{v_0}{\Omegap} \simeq 3.41 \frac{v_0}{\Omegap} \,. \label{eq:OLR}
\end{align}
The higher multipoles are
\begin{equation}
\Phi_{2n}(R) = \frac{v_0^2}{n}\left[ \frac{1-q}{1+q}\right]^n\,,
\end{equation}
where $n=1$ is the quadrupole, $n=2$ is the octupole, etc. They are all constant.

\end{document}